\documentclass[aps,twocolumn,amssymb,prl,showpacs,10pt]{revtex4-2}

\usepackage[utf8]{inputenc}
\usepackage{tabularx}
\usepackage{multirow}
\usepackage{url}
\usepackage{amsmath}
\usepackage{amssymb,amsmath}	
\usepackage{gensymb}
\usepackage{graphicx}
\usepackage{mathrsfs}
\usepackage{color}
\usepackage[normalem]{ulem}
\usepackage{float}
\usepackage{lmodern}
\usepackage{soul}
\usepackage[breaklinks,colorlinks,urlcolor=blue,citecolor=blue]{hyperref}

\newcommand{\be}{\begin{equation}}
\newcommand{\ee}{\end{equation}}
\newcommand{\bi}{\begin{itemize}}
\newcommand{\ei}{\end{itemize}}
\newcommand{\figg}[1]{Fig.~\ref{fig:#1}}
\newcommand{\eqq}[1]{Eq.~\ref{eq:#1}}

\newcommand{\ls}{}

\newcommand{\comp}{\,c/\omega_{\mathrm{p}}}

\def\ompt{\omega_{\rm p}t}

\newcommand{\apjl}{ApJ}

\begin{document}

\title{Non-ideal fields solve the injection problem in relativistic reconnection}

\author{Lorenzo Sironi}
\email{lsironi@astro.columbia.edu}
\affiliation{Department of Astronomy and Columbia Astrophysics Laboratory, Columbia University, New York, NY 10027, USA}
\date{\today}

\begin{abstract}
Magnetic reconnection in relativistic plasmas is well established as a fast and efficient particle accelerator, capable of explaining the most dramatic astrophysical flares. With particle-in-cell simulations, we demonstrate the importance of non-ideal fields for the early stages (``injection'') of particle acceleration. Most of the particles ending up with high energies (near or above the mean magnetic energy per particle) must have passed through non-ideal regions where the assumptions of ideal magnetohydrodynamics are broken (i.e., regions with $E>B$ or nonzero $E_\parallel={\bf E}\cdot {\bf B}/B$), whereas particles that do not experience non-ideal fields end up with Lorentz factors of order unity. Thus, injection by non-ideal fields is a necessary prerequisite for further acceleration. 
Our results have important implications for the origin of nonthermal particles in high-energy astrophysical sources.
\end{abstract}

\maketitle
Magnetic reconnection in the relativistic regime \citep{lyutikov_uzdensky_03,lyubarsky_05,comisso_14}, where the magnetic energy is larger than the particle rest-mass energy (equivalently, the mean magnetic energy per particle is $\sim\sigma m c^2\gg mc^2$, with $\sigma$ the magnetization), has been invoked to explain the most dramatic flaring events in astrophysical high-energy sources \citep[e.g.][]{cerutti_13a,yuan_16,lyutikov_18,petropoulou_16,nalewajko_19,christie_19,mehlhaff_20,hosking_sironi_20}. Our understanding of the physics of relativistic reconnection has greatly advanced thanks to fully-kinetic particle-in-cell (PIC) simulations, which have established reconnection as an efficient particle accelerator \citep[e.g.][]{zenitani_01,ss_14,guo_14,werner_16,zhang_sironi_21}. 
It is widely accepted that most of the energy gain of ultra-relativistic particles comes from ideal fields \cite[e.g.][]{ss_14,guo_19}. \ls{It was then argued that the spectrum of high-energy particles would remain unchanged, if non-ideal fields were to be ignored \cite{guo_19}.}

In this {\it Letter}, we 
 demonstrate that, instead, non-ideal fields have a key role in the acceleration process. They are essential for solving the ``injection problem,''  so that non-relativistic particles can be promoted to relativistic ($\sim\sigma m c^2\gg mc^2$) energies. Injection by non-ideal fields is a necessary prerequisite to access further acceleration channels, primarily governed by ideal fields \citep{ss_14,nalewajko_15,guo_19,petropoulou_18,hakobyan_21,zhang_sironi_21}. \ls{High-energy particles receive most of their energy by ideal fields (in 2D, via Fermi-like acceleration with the ``slingshot'' mechanism \cite{guo_19} or via magnetic moment conservation in the increasing field of compressing plasmoids \cite{petropoulou_18,hakobyan_21}; in 3D, via grad-B-drift acceleration while their orbits sample both sides of the layer \cite{zhang_sironi_21}).} However, we find that the particles must be pre-energized in non-ideal regions, before being further accelerated by ideal fields. 
We find that if non-ideal fields were artificially excluded, all particles would end up with low energies. 
{\it Setup}---We perform 2D and 3D PIC simulations with TRISTAN-MP \citep{buneman_93, spitkovsky_05}. We initialize a force-free field of strength $B_0$, whose direction rotates  from $+\hat{x}$ to $-\hat{x}$ across a current sheet at $y=0$. We consider a cold electron-positron plasma with \ls{rest-frame} density $n_0$ of 16 particles per cell in 2D and 4 in 3D (Suppl.~Mat. for convergence studies). Particles initially in the current sheet are excluded from our analysis, so to obtain results independent from specific choices at initialization. The field strength $B_0$ is parameterized by the magnetization $\sigma = B_0^2 / 4\pi  n_0 m c^2 = \left(\omega_{\rm c} / \omega_{\rm p}\right)^2$, where $\omega_{\rm c} = e B_0 / m c$ is the Larmor frequency and $\omega_{\rm p} = \sqrt{4\pi n_0 e^2 / m}$ the plasma frequency. We vary $\sigma$ in the range $3\lesssim\sigma\lesssim 200$, with $\sigma=50$ as our reference case. Most of our runs assume a vanishing guide field, but we also present results with guide field $B_g=B_0$ initialized along $z$ as in \citep{kilian_20}.

Along the $y$-direction of inflows, two injectors continuously introduce fresh plasma and magnetic flux into the domain, see \citep{sironi_16}. We employ periodic boundary conditions in $z$. Most of our runs have periodic boundaries also in $x$ 
(but our conclusions also hold for outflow $x$-boundaries, Suppl.~Mat.). 
Each simulation is evolved for $\sim 2.5\,L_x/c$ (the periodic  $x$-boundaries would artificially choke reconnection at  longer times). We usually let reconnection start spontaneously from numerical noise, but we also show similar results when reconnection is ``triggered'' by hand at the initial time.
 We resolve the plasma skin depth with $\comp=5$~cells, and employ large domains up to $L_x\simeq 6400\comp$ in 2D (our reference is $L_x\simeq 1600\comp$), and $L_x=2L_z\simeq 800\comp$ in 3D. 

\begin{figure*}
\centering
    \includegraphics[width=\textwidth]{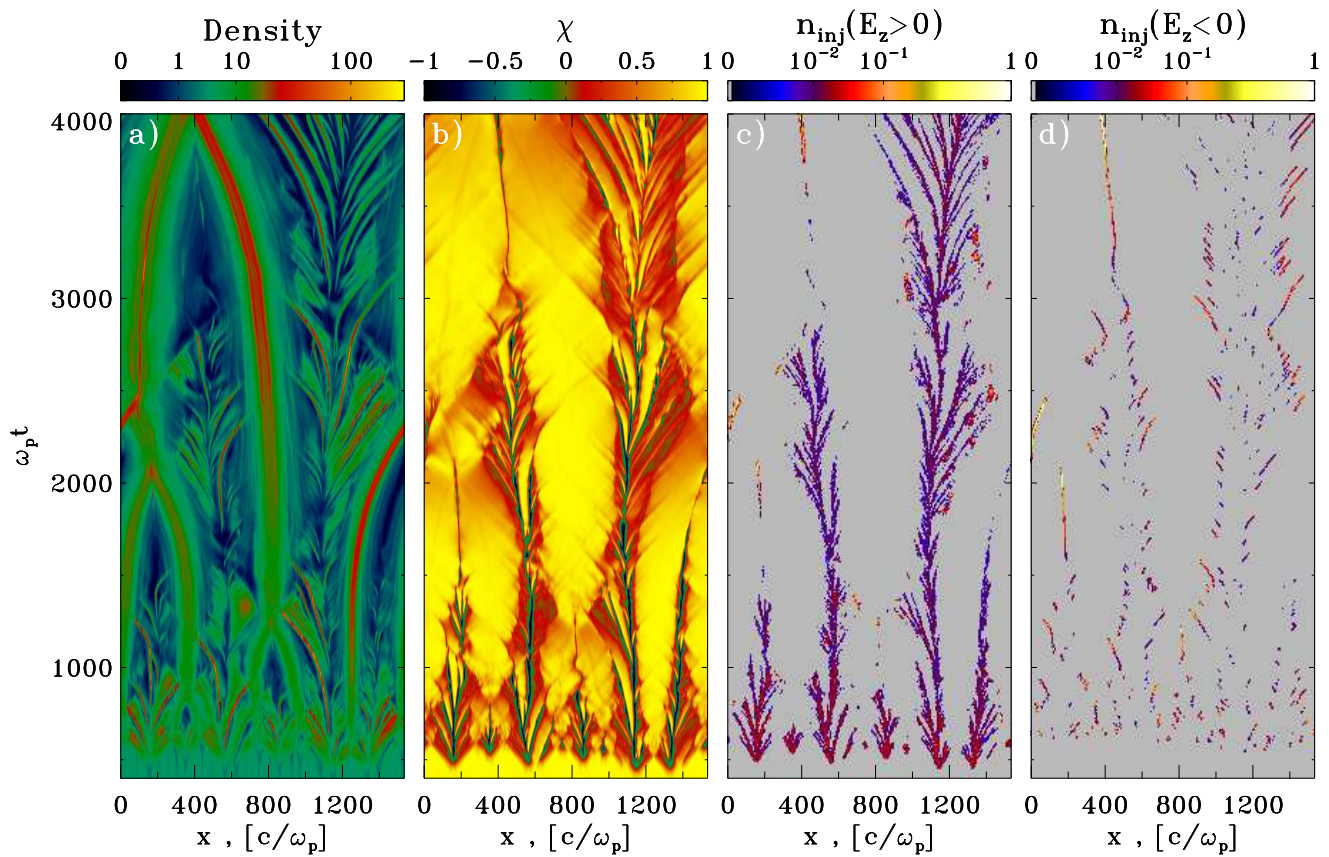}
    \caption{Evolution of the layer for our reference simulation ($\sigma=50,\;B_g=0,\;L_x=1600\comp$). (a) Density in the midplane $y=0$, normalized to $n_0$; (b) $\chi=(B^2-E^2)/(B^2+E^2)$ at $y=0$, to identify regions of magnetic ($\chi>0$) or electric ($\chi<0$) dominance; the particle $x$-locations at their first $E>B$ encounter are shown in (c) and (d), where we distinguish between particles that experience $E>B$ with $E_z>0$ (c) vs $E_z<0$ (d) (see text).
    } 
    \label{fig:fluid}
\end{figure*}

{\it Results}---Near X-points the reconnected field scales as $B_y\sim B_0\,x/\Delta$ over some length $\Delta$, whereas the electric field is $E_z\sim \eta_{\rm rec} B_0$, with $\eta_{\rm rec}\sim 0.1$ the reconnection rate \citep[e.g.,][]{sironi_16}. For $B_g=0$, magnetic dominance is broken (i.e., $E>B$) at $|x|\lesssim \eta_{\rm rec} \Delta$. For $B_g=0$ the parallel electric field $E_\parallel=\boldsymbol{E}\cdot \boldsymbol{B}/B$ necessarily vanishes,  so non-ideal effects  are captured by $E>B$, rather than by $E_\parallel$. In contrast, in the presence of a nonzero $B_g$, the region of electric dominance disappears if $B_g/B_0\gtrsim \eta_{\rm rec}$. Here, $E_\parallel\sim \eta_{\rm rec} B_0$ near the X-point, i.e., the electric field is entirely accounted for by $E_{\parallel}$. Thus, non-ideal effects are captured by $E>B$ for $B_g/B_0\lesssim \eta_{\rm rec}$, and by $E_\parallel$ for stronger guide fields. We verified this argument with a sweep of $B_g/B_0$ values, but here we focus only on $B_g=0$ (our main case) and $B_g/B_0=1$.

The layer evolution for our reference simulation ($\sigma=50,\;B_g=0,\;L_x=1600\comp$) is shown in \figg{fluid}(a), where we present the spatio-temporal structure of density in the midplane ($y=0$). At early times, the  layer
breaks into a series of primary plasmoids. Over time, they grow and coalesce, and new secondary plasmoids appear in the under-dense regions in-between primary plasmoids (e.g., top right in \figg{fluid}(a)). Regions with $E>B$ are rather ubiquitous in between plasmoids (see the green and blue areas in \figg{fluid}(b), where we plot $\chi=(B^2-E^2)/(B^2+E^2)$ at $y=0$). For each simulation particle, we detect the first time (if any) it experiences $E>B$, and record its  position at this time. The particle $x$-locations at their first $E>B$ encounter are shown in panels (c) and (d), where we distinguish between particles that experience $E>B$ with $E_z>0$ (c) vs $E_z<0$ (d). The former ($E_z>0$) is expected for X-points of the main layer, whereas $E_z<0$  in between merging plasmoids, as demonstrated by comparing (c) and (d) with (a).

\begin{figure}
\centering
    \includegraphics[width=0.48\textwidth]{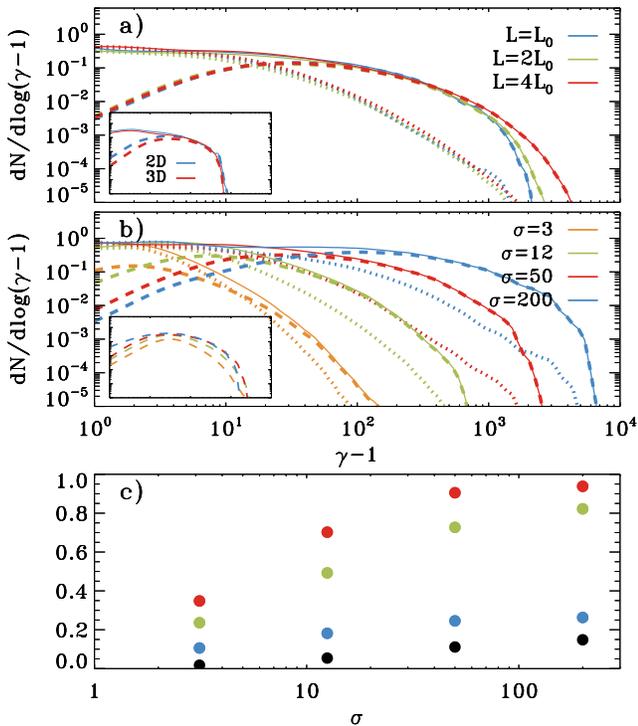}
    \caption{(a) Particle spectra at $t\simeq 1.8 \, L_x/c$ (solid lines), for simulations with $\sigma=50,\;B_g=0$ and different box sizes: $L_x=L_0=1600\comp$ (blue), $L_x=2L_0=3200\comp$ (green), and $L_x=4L_0=6400\comp$ (red). Dashed lines show the spectrum  $E>B$~particles (see text), dotted lines of $E<B$ particles. Inset: comparison of 2D and 3D triggered simulations with $\sigma=50,\;B_g=0,\;L_x=400\comp$ at $t\simeq1,000\,\omega_{\rm p}^{-1}\simeq 2.5 L_x/c$. Dashed and solid lines as in the main panel. (b) Spectra at $t\simeq 4000\, \omega_{\rm p}^{-1}\simeq 2.5 L_x/c$ for simulations with $L_x=1600\comp, \;B_g=0$ and varying $\sigma$, see the legend; line style as in (a). The inset presents the spectra of $E>B$~particles, but $\gamma-1$ on the horizontal axis is rescaled by $(50/\sigma)$. (c) Contribution of $E>B$~particles to the total census in the reconnection region (blue), and to the number of particles with $\gamma>\sigma/4$ (green) and $\gamma>\sigma$ (red), as a function of $\sigma$. The black points show the fraction of length along $y=0$ occupied by $E>B$ regions.}
    \label{fig:speccomp}
\end{figure}

The high-energy part of the spectrum extracted from the reconnection region \footnote{We define the reconnection region such that it contains a mixture of particles starting from $y>0$ and $y<0$ \citep{rowan_17}, with both populations contributing at least 10\% (our results do not significantly depend on this fraction). Unless otherwise noted, we only show the spectrum of particles belonging to the reconnection region.} is dominated by particles that experienced $E>B$ at some point in their history. In \figg{speccomp}, we plot the overall spectrum with solid lines, the spectrum of particles that experienced $E>B$ (hereafter, ``$E>B$~particles'') with dashed lines, and the spectrum of particles that never experienced $E>B$ (hereafter, ``$E<B$ particles'') with dotted lines. ``$E>B$~particles'' are particles that at some point experienced $E>B$, so they are not only those currently in $E>B$ regions. 

For high magnetizations, the high-energy tail of the spectrum is mostly populated by $E>B$~particles, regardless of system size (\figg{speccomp}(a) \footnote{The shift of the spectral cutoff to higher energies with increasing system size has been extensively characterized, in both 2D \cite{werner_16,petropoulou_18,hakobyan_21} and 3D \cite{zhang_sironi_21}.}) and dimensionality (inset in \figg{speccomp}(a), showing a comparison between 2D and 3D).
With increasing $\sigma$ (\figg{speccomp}(b)), the spectrum of $E>B$~particles shifts to higher energies as $\propto \sigma$ (see inset in \figg{speccomp}(b), where $\gamma-1$ on the horizontal axis is rescaled by $(50/\sigma)$), and it increases in normalization. In contrast, the spectrum of $E<B$ particles peaks at $\gamma-1\sim {\rm few}$ for all magnetizations, \ls{and at high energies it drops much steeper than the $E>B$ spectrum.} Thus, the overall spectrum can be described as a combination of two populations: a low-energy peak at trans-relativistic energies contributed by $E<B$ particles, and a high-energy bump with mean Lorentz factor $\propto \sigma$ populated by $E>B$~particles. 


A robust result of PIC simulations is that higher magnetizations display harder spectra, with $p=-d\log N/d\log\gamma\rightarrow 1$ for $\sigma\gg1$ \citep[e.g.][]{zenitani_01,ss_14,guo_14,werner_16}. For domain sizes within the reach of current PIC simulations, the spectrum does not extend much beyond the post-injection spectrum (the cutoff is at $\lesssim 10\,\sigma m c^2$ \cite{petropoulou_18,hakobyan_21,zhang_sironi_21}; indeed, hard power-laws with $p<2$ could not extend to much higher energies without running into an energy crisis \cite{petropoulou_18}). The fact that higher magnetizations display harder spectra  has a simple explanation.
 While the peak of the $E<B$~population is nearly $\sigma$-independent, the $E>B$~component shifts to higher energies ($\sim \sigma mc^2$) and higher normalizations with increasing $\sigma$, thus hardening the overall spectrum (\figg{speccomp}(b)). 
 
 In the asymptotic limit $\sigma\gg1$, $E>B$~particles contribute a fraction $\gtrsim 90\%$ at $\gamma>\sigma$ (red points in \figg{speccomp}(c)) and $\gtrsim 70\%$ at $\gamma>\sigma/4$ (green). These fractions are nearly constant in time for $t\gtrsim L_x/c$ (Suppl.~Mat.), and independent of the domain size (\figg{speccomp}(a)).
For $\sigma\gtrsim 50,\;E>B$~particles account for  $\sim 20\%$ of the overall census in the reconnection region (\figg{speccomp}(c), blue). 
This can be related to the fraction of length along the $y=0$ line (of area in the $y=0$ plane, for 3D) occupied by non-ideal regions, as we now explain.

The black points in \figg{speccomp}(c) denote the ``occupation fraction'' of $E>B$  regions along $y=0$, averaged over $1\lesssim ct/L_x\lesssim 2$. This fraction
increases with $\sigma$ \footnote{The increase with $\sigma$ has two reasons: first, fragmentation of the layer by the secondary plasmoid instability \citep{uzdensky_10} is more pronounced for higher $\sigma$, increasing the number of non-ideal regions \citep{sironi_16}; second, a given $E>B$ region is more extended at higher $\sigma$, since $\eta_{\rm rec}$ increases with magnetization \cite{liu_15}. Both effects reach an asymptotic limit at $\sigma\gtrsim 50$}, and in the limit $\sigma\gg1$ approaches $\sim 10\%$ (black points in \figg{speccomp}(c)), which is about half of the fraction of $E>B$~particles (blue). 

\begin{figure}
\centering
    \includegraphics[width=0.48\textwidth]{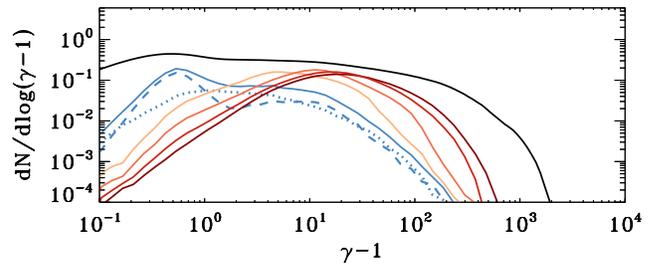}
    \caption{Evidence of fast energization near non-ideal regions, for $\sigma=50,\;B_g=0$ and $L_x=1600\comp$. The  spectrum of particles at their first $E>B$ encounter is shown by the dashed blue line if $E_z>0$, and dotted blue if $E_z<0$. Their sum is the solid blue. The series of spectra from light to dark red are measured, for those same particles, respectively  $\sim 9,\, 27,\, 90,\, 270\;\omega_{\rm p}^{-1}$ after their first $E>B$ encounter. For comparison, the black line shows the overall spectrum at the final time $\ompt\simeq 4000$.}
    \label{fig:specx}
\end{figure}

This factor of two has a simple explanation. The dashed blue line in \figg{specx} shows the spectrum measured, for each particle, at its first $E>B$ encounter with $E_z>0$, as appropriate for X-points in the main layer. It contains $\sim 10\%$ of post-reconnection particles, i.e., exactly equal to the $E>B$ occupation fraction.
The dotted blue line in \figg{specx}, instead, shows the spectrum of particles experiencing $E>B$ with $E_z<0$, i.e., in between merging plasmoids. It also contains $\sim 10\%$ of particles. Thus, for $\sigma\gtrsim 50$, $\sim 10\%$ of particles encounter $E>B$ fields when entering the reconnection region, and an additional $\sim 10\%$ in secondary layers between merging plasmoids. The latter extend along $y$, so their X-points are not accounted for by the black markers in \figg{speccomp}(c). This justifies why for $\sigma\gtrsim 50$ the fraction of $E>B$~particles (blue in \figg{speccomp}(c)) is twice larger than the $E>B$ occupation fraction (black).


In \figg{specx}, we provide evidence of fast particle acceleration near non-ideal regions. The  spectrum of particles at their first $E>B$ encounter is shown by the solid blue line, demonstrating that at this point 
the particles still have low energies.
The series of spectra from light to dark red are measured, for those same particles, respectively  $\sim 9,\, 27,\, 90,\, 270\,\omega_{\rm p}^{-1}$ after their first $E>B$ encounter. The spectral peak quickly shifts up to $\gamma-1\sim 5$ (first red line; at this time, most of the particles are still in $E>B$ regions), yielding a mean acceleration rate $d\gamma/dt\sim 0.5\,\omega_{\rm p}^{-1}$, comparable to the maximal rate $\sim \eta_{\rm rec}|\beta_z|\sqrt{\sigma}\omega_{\rm p}^{-1}$ \cite{zhang_sironi_21} assuming a $z$-velocity $|\beta_z|\simeq0.7$ (particles accelerated at X-points also have some $\beta_x$ along the outflow). 
Rapid acceleration continues up to $\gamma-1\simeq 20\sim \sigma/2$ (third red line) \footnote{As we show in Suppl.~Mat., this is also the peak of the spectrum of particles currently residing in $E>B$ regions.}. Beyond this stage, the spectrum still shifts up in energy  \cite{petropoulou_18,hakobyan_21}, but at a slower rate (compare the two darkest red lines). The later stages are governed by ideal fields \cite{guo_19}.

\begin{figure}
\centering
    \includegraphics[width=0.48\textwidth]{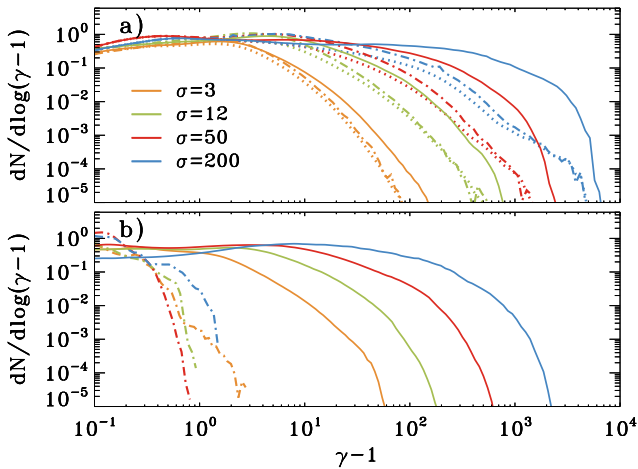}
    \caption{Experiments with test-particles that do not experience non-ideal fields, for $B_g=0$  (top) and $B_g=B_0$ (bottom). Spectra are computed at $\ompt=4000$ for simulations with $L_x=1600\comp$ and varying $\sigma$ (see legend). Top: solid and dotted lines as in \figg{speccomp}(b), whereas dash-dotted lines present the final spectrum of test-particles. They are initialized as regular particles, but when they pass through $E>B$ regions, we artificially fix their Lorentz factor at $\gamma-1\sim {\rm few}$ (more specifically, to the same peak as the dotted lines). Bottom: solid lines for regular particles, while dash-dotted lines for test-particles evolved without the contribution by $E_\parallel$. }
    \label{fig:specchi}
\end{figure}

To corroborate our conclusions on the importance of non-ideal fields for particle injection, we present in \figg{specchi} two additional experiments. For $3\lesssim \sigma\lesssim 200$, we analyze the cases of vanishing ($B_g=0$, top) and strong ($B_g=B_0$, bottom) guide fields. The final spectra are indicated by solid lines. In the top panel, dotted lines show the spectra of $E<B$~particles, whereas dash-dotted lines the spectra of ``test-particles'' --- not contributing to the electric currents in the simulation, but otherwise initialized and evolved as regular particles. When test-particles pass through $E>B$ regions, we artificially fix their Lorentz factor at $\gamma-1\sim {\rm few}$. The remarkable agreement between dash-dotted and dotted lines demonstrates that if we do not allow  test-particles to gain energy while in $E>B$ regions, they display similar spectra as particles that never had $E>B$ encounters (we have confirmed this  also for the 3D simulation of \figg{speccomp}(a), inset). Equivalently, energization in non-ideal regions plays a key role in shaping the high-energy end of the particle spectrum.

\figg{specchi}(b) refers instead to $B_g=B_0$. As we discussed, here non-ideal effects are well captured by $E_\parallel$. {Comparison with \figg{specchi}(a)  shows that spectra are softer for larger $B_g$. The fraction of injected particles decreases with increasing $B_g/B_0$ because the layer is less prone to fragmentation into plasmoids \citep{werner_17,ball_19}, and so to formation of non-ideal regions.} 
In \figg{specchi}(b), dash-dotted lines present the spectrum of test-particles evolved without $E_\parallel$, so in response to $\boldsymbol{E}-E_\parallel(\boldsymbol{B}/B)$. 
When inhibiting energization by $E_\parallel$, the test-particles stay at non-relativistic energies.

{\it Conclusions}---We investigate the injection physics of particle acceleration in relativistic reconnection. In contrast to earlier claims \cite{guo_19}, we find that energization by non-ideal fields is a necessary prerequisite for further acceleration (in Suppl.~Mat. we compare to \cite{guo_19}). Particles that are artificially evolved without non-ideal fields do not even reach relativistic energies. \ls{While it is true that high-energy particles receive most of their energy via ideal fields \cite{guo_19}, this can only happen after an injection phase necessarily governed by non-ideal fields. We then argue that studies of reconnection-powered acceleration that employ test-particles in magnetohydrodynamics simulations need to properly include non-ideal fields.}

The spectral component of particles that encountered non-ideal regions shifts to greater energies ($\sim \sigma m c^2$) and higher normalizations  with increasing magnetization,  whereas particles that do not experience non-ideal fields always end up with Lorentz factors near unity. The overall spectrum then gets harder for higher $\sigma$, which explains 
the $\sigma$-dependent spectral hardness reported in PIC simulations  \citep[e.g.,][]{ss_14,guo_14,werner_16}. This statement applies to the range $1\lesssim \gamma\lesssim 10\,\sigma$ of the post-injection spectrum.
At higher energies, an additional power-law tail will emerge, whose slope is set by the dominant acceleration mechanism, which is different between 2D \cite{uzdensky_20,hakobyan_21} and 3D \cite{zhang_sironi_21}.

Finally, we remark that, even though we have assumed an electron-positron plasma, it is well known that $\sigma\gg1$ reconnection behaves similarly in electron-positron, electron-proton \citep{guo_16b,werner_18,ball_18} and electron-positron-proton \citep{petropoulou_19} plasmas, so our results should apply regardless of the plasma composition. \ls{The importance of non-ideal fields for particle injection has also been emphasized in studies of non-relativistic low-beta turbulence and reconnection \cite{blackman_94,dmitruk_04,dahlin_14,dalena_14} and magnetically-dominated turbulence \citep{comisso_sironi_19}.}


\phantom{xx}
\begin{acknowledgments}
We thank L. Comisso, D. Groselj, A. Spitkovsky and N. Sridhar for comments that greatly improved the clarity of the paper. We thank F. Guo for discussions on this topic. L.S. acknowledges support from the Cottrell Scholars Award, NASA 80NSSC20K1556, NSF PHY-1903412, DoE DE-SC0021254 and NSF AST-2108201.  This project made use of the following computational resources: NASA Pleiades supercomputer, Habanero and Terremoto HPC clusters at Columbia University. This work is dedicated to the memory of my father.
\end{acknowledgments}

\section{Supplementary Material}
\subsection{Secondary Particle Energization}
\begin{figure}
\centering
    \includegraphics[width=0.48\textwidth]{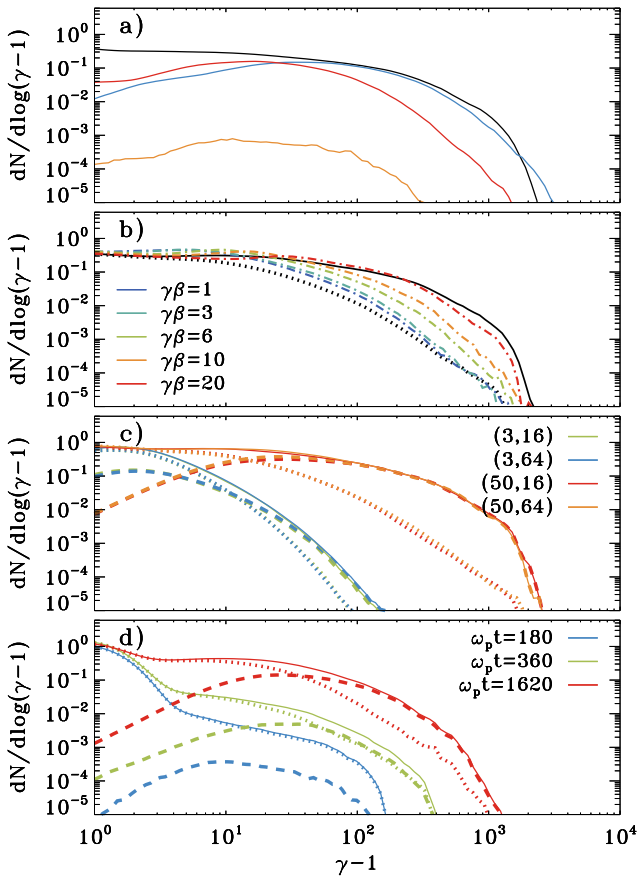}
    \caption{(a) In orange, we show a representative spectrum of particles that currently reside in $E>B$ regions, measured at $\ompt\simeq900$ for our reference simulation with $\sigma=50$ and $L_x=1600\comp$. The red line is obtained by integrating over time (until the final time $\ompt\simeq 4000$) the instantaneous spectra of $E>B$ regions, assuming that particles spend there an average time of $t\sim 10\,\omega_{\rm p}^{-1}$. The blue line is obtained from the $E>B$ spectra measured at 450 different time snapshots (one every $9\,\omega_{\rm p}^{-1}$), by extrapolating the particle energies to later times as discussed in the text (see \eqq{eq1}).
    For comparison, the black line shows the overall spectrum in the reconnection region at the final time ($\ompt\simeq 4000$). (b) The solid black line is the total particle spectrum at the final time ($\ompt=4000$) for our fiducial simulation with $\sigma=50$ and $L_x=1600\comp$. The dotted black line shows the spectrum of $E<B$ particles at the same time. Dash-dotted lines present the final spectrum of test-particles, which are initialized as regular particles, but when they pass through $E>B$ regions we artificially fix their $\gamma\beta$ to the value indicated in the legend. (c) Convergence with respect to the number of particles per cell, which we vary from 16 (green and red) to 64 (blue and orange), for two magnetizations, as indicated in the legend: $\sigma=3$ (green and blue) and $\sigma=50$ (red and orange). The simulations have $L_x=1600\comp$. (d) Comparison with \citep{guo_19} (see the text for details), showing the particle spectra at three different times, as indicated in the legend. In both panels (c) and (d), we show the total particle spectrum (solid), as well as the spectrum of $E>B$ particles (dashed) and of $E<B$ particles (dotted). All the spectra in this figure refer to $B_g=0$ simulations. }
    \label{fig:specppc}
\end{figure}

In \figg{specppc}(a) we present with the orange line a representative spectrum of particles that  currently reside in $E>B$ regions. The spectrum is measured at $\ompt\simeq900$ for our reference simulation with $\sigma=50$, $B_g=0$ and $L_x=1600\comp$. The red line, instead, is obtained by integrating over time (until the final time $\ompt\simeq 4000$) the instantaneous spectra from $E>B$ regions, assuming that particles spend there an average time of $\sim 10\,\omega_{\rm p}^{-1}$. The red spectrum has a broad peak at $\gamma-1\sim 20
$, i.e., the characteristic Lorentz factor of particles in $E>B$ regions is $\gamma-1\sim20\sim \sigma/2$.
For comparison, the black line shows the overall spectrum in the reconnection region at the final time. 

The blue spectrum is computed as follows: for each of the 450 output snapshots of our simulation, we consider the instantaneous (at time $t_i$) spectrum of particles located in $E>B$ regions, and we extrapolate their energy to the final time $t_f\simeq 4000\,\omega_{\rm p}^{-1}$ as
\be\label{eq:eq1}
\gamma_f-1=(\gamma_i-1) \left(1+\frac{t_f-t_i}{\Delta t}\right)^{1/2}
\ee
where $\gamma_i$ is the Lorentz factor at the injection time $t_i$ and $\gamma_f$ the one at $t_f$, and we have chosen $\Delta t=0.2\,L_x/c\simeq300\,\omega_{\rm p}^{-1}$. The blue line is then obtained by summing the contributions of the individual snapshots $t_i$, still assuming that particles spend in $E>B$ regions an average time of $t\sim 10\,\omega_{\rm p}^{-1}$ (as we assumed for the red line).

The scaling in \eqq{eq1} is  motivated by the dominant mechanism of secondary particle energization in 2D relativistic reconnection \citep{petropoulou_18,hakobyan_21} --- beyond the primary acceleration/injection discussed in this paper. 2D PIC simulations have shown that the Lorentz factor of high-energy particles scales as $\gamma\propto t^{1/2}$. This is driven by a linear increase in the  field strength felt by the particles trapped in plasmoids, coupled with the conservation of their first adiabatic invariant \citep{petropoulou_18,hakobyan_21}. The blue line is in good agreement --- as regard to both shape and normalization --- with the high-energy part of the overall spectrum (black). This demonstrates that primary acceleration/injection at/near $E>B$ regions, followed by secondary acceleration in compressing plasmoids (as prescribed by \eqq{eq1}, and appropriate for 2D), provides an excellent description of the history of high-energy particles. In particular, even though at any given time the fraction of particles residing in $E>B$ regions is small (orange line), their time-integrated contribution (blue line) fully accounts for the high-energy part ($\gamma\gtrsim \sigma$) of the final spectrum (black line).

\section{Test-particles in $B_g=0$ simulations}
 In \figg{specppc}(b), we show with the solid black line the total particle spectrum at the final time ($\ompt=4000$) for our fiducial simulation with $\sigma=50$, $B_g=0$ and $L_x=1600\comp$. The dotted black line shows the spectrum of $E<B$ particles at the same time, whereas dash-dotted lines present the final spectrum of test-particles. They are initialized as regular particles, but when they pass through $E>B$ regions, we artificially fix their $\gamma\beta=\sqrt{\gamma^2-1}$ to the value indicated in the legend. The dash-dotted lines for $\gamma\beta=1$ and $3$ are nearly identical, demonstrating that our conclusions are robust as long as test-particles are constrained to stay at trans-relativistic energies while in $E>B$ regions. Both cases are similar to the dotted black line, i.e., test-particles that are forced to keep a low $\gamma\beta$ while in $E>B$ regions display similar spectra as particles that never had $E>B$ encounters. 
 
 Higher values of $\gamma\beta$ bring the high-energy end of the test-particle spectrum closer to the one of regular particles (black solid line). In particular, the spectrum of $\gamma\beta=20$ test-particles nearly overlaps with the spectrum of regular particles. Once again, this suggests that $E>B$ regions lead to acceleration up to $\gamma\sim 20\sim \sigma/2$, as demonstrated by Fig.~3 in the main paper. As long as $E>B$ regions are allowed (or constrained, as we do here with test-particles) to accelerate particles up to $\gamma\sim \sigma/2$, the resulting test-particle spectrum is consistent with the spectrum of regular particles.

\subsection{Convergence studies}
In \figg{specppc}(c), we show that the total particle spectra (solid lines), as well as the spectra of $E>B$ particles (dashed lines) and of $E<B$ particles (dotted lines), are identical when employing 16 particles per cell (our fiducial value) or 64 particles per cell. This conclusion holds for both low ($\sigma=3$) and high ($\sigma=50$) magnetizations.

\subsection{Comparison with Guo et al. (2019)}
In \figg{specppc}(d), we directly compare our results to the work by \citep{guo_19}, since their conclusions are in contradiction with our findings. \ls{In particular, they reported evidence that most high-energy particles never had $E>B$ encounters, and they argued that the spectrum of high-energy particles would remain the same if non-ideal fields were to be excluded.}

To ensure a fair comparison, we initialize the plasma with the same temperature as in \citep{guo_19} ($kT=0.36\,m c^2$) and we initiate reconnection by hand (``triggered'' setup, where we reduce the pressure near the center $(x,y)=(0,0)$ of the layer at the initial time). Our spatial resolution $\comp=5$ is larger than in \citep{guo_19}, and our box length $L_x=1600\comp$ is also greater. The ``cold'' magnetization (normalized to the plasma rest mass energy density) is $\sigma=50$, corresponding to an effective magnetization (normalized to the enthalpy density) of $\sigma_{\rm eff}\simeq 24$. The simulation has $B_g=0$.

In all the spectra showed so far, we only accounted for particles belonging to the reconnection region (defined such that it contains a mixture of particles starting from $y>0$ and $y<0$ \citep{rowan_17}, with both populations contributing at least 10\%). Instead, in order to compare directly with Fig.~3(c) by \citep{guo_19}, \figg{specppc}(d) includes all particles within $\sim 0.25 L_x$ from the midplane $y=0$ (upstream particles populate the low-energy bump at $\gamma-1\lesssim 3$). We present the total  particle spectrum (solid), as well as the spectrum of $E>B$ particles (dashed) and of $E<B$ particles (dotted), at three different times, as indicated by the legend. 
While at early times the high-energy end is mostly populated by $E<B$ particles, at later times --- when the spectrum has finally developed a significant high-energy component --- the situation is reversed, with $E>B$ particles clearly dominating the high-energy spectrum. At late times, the total particle spectrum (solid)  is significantly harder than the spectrum of $E<B$ particles (dotted). \ls{We therefore disagree with the conclusions by \citep{guo_19}, who claimed that most high-energy particles never had an $E>B$ encounter, and that the spectrum of high-energy particles would remain the same if non-ideal fields were to be ignored.}

Note that the trend we observe, with $E>B$ particles becoming dominant at later times, can also be seen in the time sequence displayed by Fig.~3(c) of \citep{guo_19}, yet this was apparently overlooked when deriving their conclusions.


\subsection{Steady state}
\begin{figure}
\centering
    \includegraphics[width=0.48\textwidth]{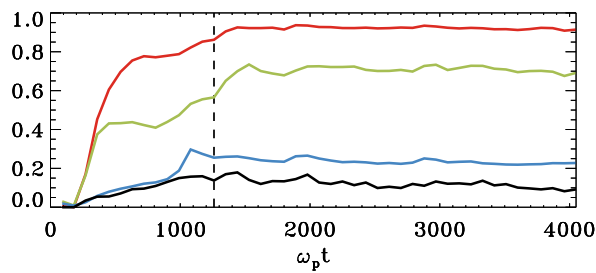}
    \caption{Time evolution of the fraction of $E>B$ particles in the reconnection region (blue), and the fraction of $E>B$ particles with $\gamma>\sigma/4$ (green) and $\gamma>\sigma$ (red). The black line shows the fraction of length along $y=0$ occupied by $E>B$ regions. We adopt the same physical parameters as our fiducial run ($\sigma=50$, $B_g=0$ and $L_x=1600\comp$), but we employ outflow boundaries in $x$ and a triggered setup. The vertical dashed line indicates the time when the two reconnection fronts that were generated at $t=0$ near the center of the domain advect out of the boundaries. This illustrates that the fractions of $E>B$ particles and the occupation fraction of $E>B$ regions attain a quasi-steady state at $\ompt\gtrsim1200$.}
    \label{fig:time}
\end{figure}

In \figg{time}, we show the time dependence of the fractions reported in Fig.~2(c) of the main paper (same color coding). We display the contribution of $E>B$ particles to the total census in the reconnection region (blue), and to the number of particles with $\gamma>\sigma/4$ (green) and $\gamma>\sigma$ (red), as a function of time. The black line shows the fraction of length along $y=0$ occupied by $E>B$ regions. To ensure that the system can achieve a statistical steady state, here we employ a triggered setup and outflow boundary conditions in the $x$ direction. Otherwise, the numerical and physical parameters are the same as in our fiducial run, with $\sigma=50$, $B_g=0$ and $L_x=1600\comp$. In contrast to the case of periodic boundaries in $x$, which artificially choke reconnection after a few $L_x/c$, with outflow boundaries one can follow the quasi-steady evolution of the system as long as computational resources allow \cite{sironi_16,zhang_sironi_21}.

\figg{time} yields two important conclusions. First, a quasi-steady state is established at $\ompt\gtrsim 1200\sim 0.75 L_x/c$. This is the time when the two reconnection fronts that were generated at $t=0$ near the center of the domain advect out of the boundaries (for details, see \cite{sironi_16}). Both the fractions of $E>B$ particles and the occupation fraction of $E>B$ regions along $y=0$ are nearly constant in time. Second, these fractions are remarkably the same for the outflow simulation presented in \figg{time} and for the periodic run in Fig.~2(c) (see the data points at $\sigma=50$). In particular, it still holds true that $E>B$ particles contribute a fraction $\sim 20\%$ of the particle census in the reconnection region, which is twice larger than the occupation fraction of $E>B$ regions. See the main text for the explanation of this factor of two.

\subsection{Particle trajectories}
\ls{We present a representative set of particle trajectories in the animation \href{http://user.astro.columbia.edu/~lsironi/particletrack.mov}{particletrack.mov}. We employ our fiducial case with $\sigma=50$, $B_g=0$ and $L_x=1600\comp$. The top panel shows the trajectory of 1500 simulation particles (yellow points), selected randomly to start around $y\simeq170\comp$ at $\ompt\simeq650$. We follow these particles over time, superimposed over the 2D plot of density (grey scale). When a particle interacts for the first time with an $E>B$ region (i.e., the particle is injected), it is depited as an open green circle (for each particle, the green circle is shown only at the time of injection). The title of the second panel tracks the fraction of particles (from the set of yellow points) that over time get injected at $E>B$ regions. The fraction of injected particles rises up to 7\% after their first interaction with the layer, and then further increases up to 18\%, primarily as a result of the merger of two large plasmoids. This is in line with what we describe in the main paper when commenting on Fig.~2(c).}

\ls{The bottom panel of the animation displays the energy history of three particles, whose spatial trajectory is shown in the top panel by the filled circles with the same color. The particles are selected randomly for each decade in Lorentz factor ($\gamma-1$ from $1$ to 10, from $10$ to 100, from 100 to 1000). The low-energy particle (yellow) belongs to the $E<B$ population. This is expected, since most particles ending up with low energies do not go through non-ideal regions. Instead, both the orange and the red particles belong to the $E>B$ population; again, this is expected, since $E>B$ particles dominate the high-energy end of the spectrum. For the orange and red particles, their injection time (i.e., the first time they experience $E>B$) is denoted by the vertical dashed line having the same color.} 

\bibliographystyle{apsrev}
\bibliography{blob.bib}
\end{document}